\documentclass[preprint,aps,floatfix,showpacs,nofootinbib]{revtex4-1}
\usepackage{ulem}
\usepackage{dcolumn}
\usepackage{bm}
\usepackage{color}
\usepackage{amssymb}
\usepackage{amsmath}
\usepackage{graphicx}
\usepackage{amsfonts}
\usepackage{slashed}
\usepackage{pstricks}
\usepackage{float}
\usepackage{hyperref}
\usepackage{array}
\allowdisplaybreaks
\begin{document}
\title{Tritium $\beta$-decay in chiral effective field theory}
\author{A. Baroni$^{\,{\rm 1}}$,
L.\ Girlanda$^{\,{\rm 2,3}}$,
A.\ Kievsky$^{\,{\rm 4}}$,
L.E.\ Marcucci$^{\,{\rm 4,5}}$,
R.\ Schiavilla$^{\,{\rm 1,6}}$,
and M.\ Viviani$^{\,{\rm 4}}$}
\affiliation{
$^{\rm 1}$\mbox{Department of Physics, Old Dominion University, Norfolk, VA 23529} \\
$^{\rm 2}$\mbox{Department of Mathematics and Physics, University of Salento, 73100 Lecce, Italy} \\
$^{\rm 3}$\mbox{INFN-Lecce, 73100 Lecce, Italy}\\
$^{\rm 4}$\mbox{INFN-Pisa, 56127 Pisa, Italy}\\
$^{\rm 5}$\mbox{Department of Physics, University of Pisa, 56127 Pisa, Italy}\\
$^{\rm 6}$\mbox{Theory Center, Jefferson Lab, Newport News, VA 23606}\\
}
\date{\today}
\begin{abstract}
We evaluate the Fermi and Gamow-Teller (GT) matrix elements in tritium
$\beta$-decay by including in the charge-changing weak current the
corrections up to one loop recently derived in nuclear chiral effective
field theory ($\chi$EFT).  The trinucleon wave functions are obtained from 
hyperspherical-harmonics solutions of the Schr\"odinger equation with
two- and three-nucleon potentials corresponding to either $\chi$EFT
(the N3LO/N2LO combination) or meson-exchange phenomenology
(the AV18/UIX combination).  We find that contributions due to loop
corrections in the axial current are, in relative terms, as large as (and
in some cases, dominate) those from one-pion exchange, which nominally
occur at lower order in the power counting.  We also provide values for
the low-energy constants multiplying the contact axial current and three-nucleon potential,
required to reproduce the experimental GT matrix element and trinucleon
binding energies in the N3LO/N2LO and AV18/UIX calculations.
\end{abstract}
\maketitle
\section{Introduction}
\label{sec:intro}
Recently, nuclear axial current and charge operators have been
derived in chiral effective field theory ($\chi$EFT) up to one loop
in a formalism based on time-ordered perturbation theory, in which, along with
irreducible contributions, non-iterative terms in reducible contributions
were identified and accounted for order-by-order in the power counting~\cite{Baroni16}.
Ultraviolet divergencies associated with the loop corrections were isolated
in dimensional regularization.  The resulting axial current was found to be finite and conserved
in the chiral limit, while the axial charge required renormalization.  In particular, the divergencies
in the loop corrections to the one-pion exchange axial charge were reabsorbed by renormalization
of some of the low-energy constants (LECs) $d_i$ characterizing the sub-leading $\pi N$
Lagrangian ${\cal L}^{(3)}_{\pi N}$~\cite{Fettes00}. For a detailed discussion of these issues (formalism,
renormalization, etc.) we defer to Ref.~\cite{Baroni16}.  However, a brief summary is provided in the
next section.
\begin{figure}[bth]
\includegraphics[width=16cm]{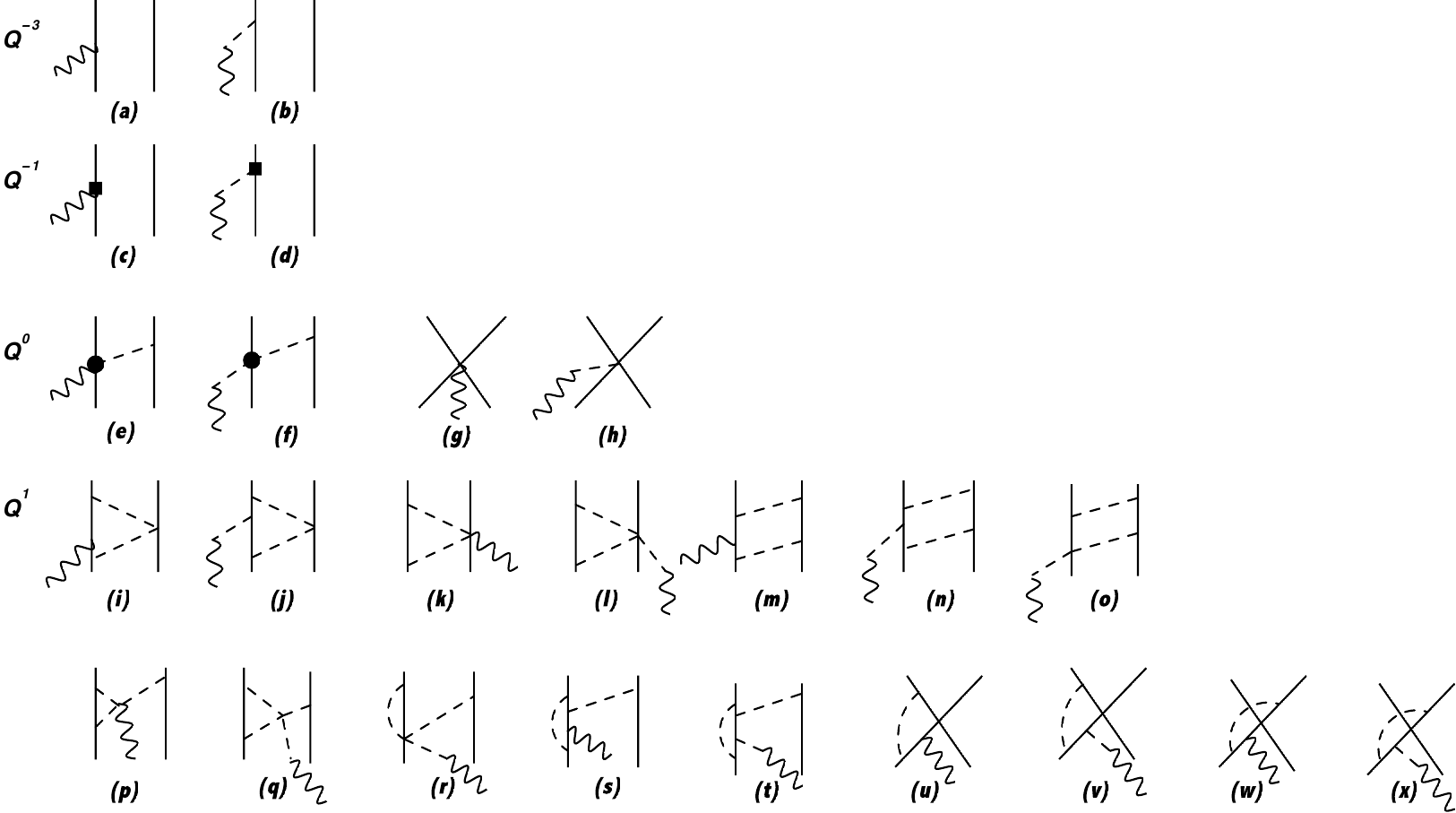}
\caption{Diagrams illustrating the one- and two-body axial currents entering at order $Q^{-3}$ (LO),
$Q^{-1}$ (N2LO), $Q^{\,0}$ (N3LO), and $Q^{\,1}$ (N4LO), where $Q$ denotes generically
the low-momentum scale.  Nucleons, pions, and axial fields
are denoted by solid, dashed, and wavy lines, respectively.  The squares in panels (c) and
(d) denote relativistic corrections to the one-body axial current, while the circles in panels
(e) and (f) represent vertices implied by the ${\cal L}^{(2)}_{\pi N}$ chiral Lagrangian, involving
the LECs $c_i$ (see Ref.~\cite{Baroni16} for additional explanations).  Only a single time
ordering is shown; in particular, all direct- and crossed-box diagrams are accounted for.  The
contributions associated with diagrams (w) and (x) were overlooked in Ref.~\cite{Baroni16}.}
\label{fig:f1}
\end{figure}

In the present paper, the focus is on the axial current, whose contributions  up to one loop
are illustrated diagrammatically in Fig.~\ref{fig:f1}.  Pion-pole terms are crucial for the current
to be conserved in the chiral limit~\cite{Baroni16}---these terms were ignored in the earlier
studies of Park {\it et al.}~\cite{Park93,Park03}; of course, they are suppressed in low momentum
transfer processes such as the tritium $\beta$-decay under consideration here. Vertices involving
three or four pions, such as those, for example, occurring in panels (l), (p), (q), and (r) of Fig.~\ref{fig:f1},
depend on the pion field parametrization.  This dependence must cancel out after summing the
individual contributions associated with these diagrams, as indeed it does~\cite{Baroni16} (this
and the requirement that the axial current be conserved in the chiral limit provide useful checks
of the calculation).

In Fig.~\ref{fig:f1} the labeling N$n$LO corresponds to the power counting
$Q^n \times Q^{\, {\rm LO}}$, where $Q$ denotes generically the low momentum
scale and $Q^{\, {\rm LO}}$ is $Q^{-3}$ for the axial
current~\cite{Baroni16}. The LO and N2LO currents consist of single-nucleon
terms; the N2LO current includes relativistic corrections proportional to
$1/m^2$ ($m$ is the nucleon mass), suppressed by two powers of $Q$
relative to the LO.  Pion-range currents contribute at N3LO, panels (e)
and (f) of Fig.~\ref{fig:f1}, and involve vertices from the sub-leading
${\cal L}^{(2)}_{\pi N}$ chiral Lagrangian~\cite{Fettes00}, proportional
to the LECs $c_3$, $c_4$, and $c_6$.  At this order (N3LO) there is
also a contact current proportional to a single LEC, which we denote as $z_0$
following Ref.~\cite{Baroni16}. This LEC is related to
the LEC $c_D$ (in standard notation), which enters the three-nucleon chiral
potential at leading order.  The two LECs $c_D$ and $c_E$ which fully
characterize this potential have recently been constrained by reproducing
the empirical value of the Gamow-Teller (GT) matrix element in tritium $\beta$
decay and the binding energies of the trinucleons~\cite{Gazit09,Marcucci12}.
However, the value determined for $z_0$ in those earlier studies was based
on calculations which retained only terms up to N3LO in the axial current.
As a matter of fact, one of the goals of the present work is to provide a determination
of $z_0$ by also accounting for the N4LO corrections, represented by diagrams
(i)-(x) in Fig.~\ref{fig:f1}. 

Most calculations of nuclear axial current matrix elements, such as those
reported for the $pp$ and $p\,^3$He weak fusions of interest in solar physics
in Refs.~\cite{Park03,Marcucci13}, and for muon capture on $^2$H and
$^3$He in Ref.~\cite{Marcucci12}, have ignored these N4LO corrections.  One
exception is Ref.~\cite{Klos13}, which included effective one-body reductions, for use
in a shell-model study, of some of the two-pion exchange terms derived
in Ref.~\cite{Park03}.  However, a systematic study of axial current
contributions at N4LO is still lacking. The other goal of the present work is to
provide a numerically exact estimate of these contributions in the $^3$H
GT matrix element.

\section{Formalism}
\begin{figure}[bth]
	\includegraphics[width=5.5in]{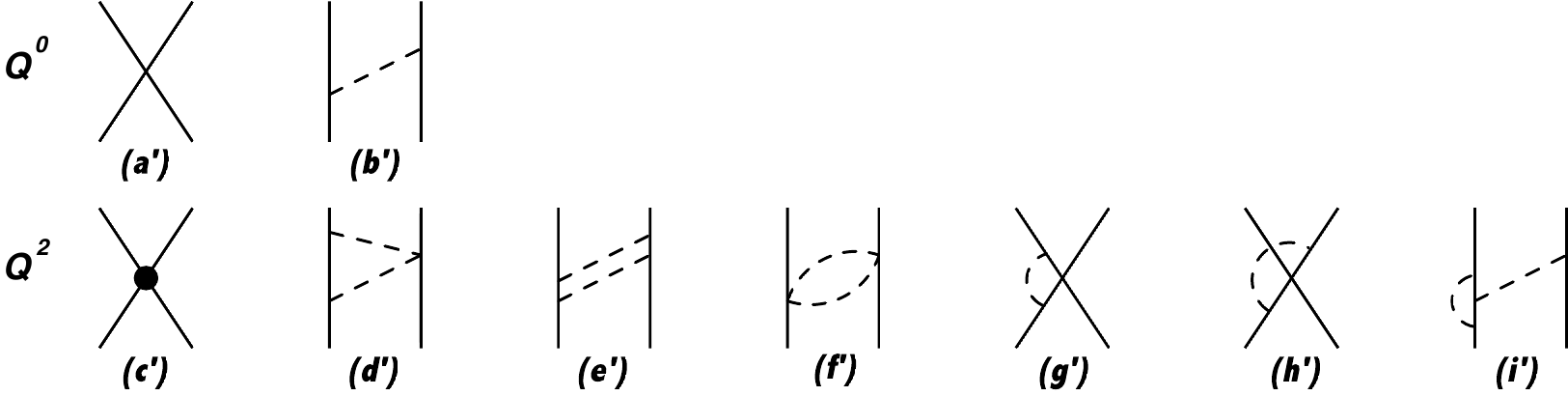}
	\caption{Diagrams illustrating contributions to the two-nucleon potential entering at $Q^0$, panels (a') and (b'), and at $Q^2$, panels (c')-(i').  Nucleons and pions are denoted
		by solid and dashed lines, respectively.  The filled circle in panel (c') represents the vertex from contact Hamiltonians containing two gradients of the nucleons' field. Vertex corrections coming from ${\cal L}_{\pi\, N}^{(3)}$ as well as $1/m$ corrections to the vertices and energy denominators, entering at order $Q^2$, are not displayed. Only a single time ordering
        for each topology is shown. In particular all direct- and crossed-box diagrams are accounted for.}
	\label{fig:fpot}
\end{figure}

The starting point of the derivation of nuclear current operators is the chiral Lagrangian for interacting pions and nucleons. This defines a quantum field theory which satisfies, besides all common general properties, like unitarity, analiticity, crossing symmetry and cluster decomposition, all constraints from chiral symmetry, in the form of chiral Ward identities, e.g., (partial) current conservation. Due to the (pseudo-) Goldstone boson character of the pions, their interactions can be organized according to increasing powers of their momenta, whose magnitude is generically denoted $Q$, much smaller than the hadronic scale $\Lambda_\chi \sim 1$~GeV.
From the chiral Lagrangian one can derive, in the canonical formalism, the chiral Hamiltonian, divided into a free part $H_0$ and an interacting part $H_I$, which allows one to calculate transition amplitudes by  applying the rules of time-ordered perturbation theory (TOPT),
\begin{equation}
\langle f \!\mid T\mid\! i \rangle= 
\langle f\! \mid H_I \sum_{n=1}^\infty \left( 
\frac{1}{E_i -H_0 +i\, \eta } H_I \right)^{n-1} \mid\! i \rangle \ .
\label{eq:pt}
\end{equation}
The evaluation of this amplitude is in practice carried out by inserting
complete sets of $H_0$ eigenstates between successive terms of $H_I$.
Power counting is then used to organize the diagrammatic expansion (which in general will involve reducible---i.e., with purely nucleonic intermediate states---and irreducible contributions) in powers of $(Q/\Lambda_\chi)\ll 1$.  In this expansion we also take into account
non-static contributions which represent nucleon-recoil corrections, by expanding a generic energy denominator as
\begin{equation}
\frac{1}{E_i-E_I-\omega_\pi}= -\frac{1}{\omega_\pi}
\bigg[ 1 + \frac{E_i-E_I}{\omega_\pi}+
\frac{(E_i-E_I)^2}{\omega^2_\pi} + \dots\bigg] \ ,
\label{eq:deno}
\end{equation}
where $E_I$ denotes the kinetic energy of the intermediate purely-nucleonic state, $\omega_\pi$ the pion energy (or energies, as the case may be), and the
ratio $(E_i-E_I)/\omega_\pi$ is of order $Q$. 
As a result the scattering amplitude $T$ admits the following expansion:
\begin{equation}
T=T^{(n)} + T^{n+1)} + T^{(n+2)} + \dots \ ,
\label{eq:tmae}
\end{equation}
where $T^{(m)} \sim Q^m$, and chiral symmetry ensures that
$n$ is finite. In the case of the two-nucleon amplitude $n=0$.  
%Divergent contributions coming from loop integrations are handled in dimensional regularization and absorbed by the process of renormalization.
Obviously, an infinite set of contributions to the TOPT expansion must be resummed in order to describe nuclear bound states. This is achieved by definining a kernel that satisfies a Lippmann-Schwinger (LS) equation and generates the above perturbative expansion of the scattering amplitude. Thus, a two-nucleon potential $v$ can be derived, assumed to admit the same kind of low-energy expansion as in Eq.~(\ref{eq:tmae}), which when iterated in the LS equation,
\begin{equation}
v+v\, G_0\, v+v\, G_0 \, v\, G_0 \, v +\dots \ ,
\label{eq:lse}
\end{equation}
where $G_0$ denotes the free two-nucleon propagator $G_0=1/(E_i-E_I+i\, \eta)$, leads to the on-the-energy-shell ($E_i=E_f$) $T$-matrix in Eq.~(\ref{eq:tmae}),
up to any specified order in the power counting.  In this way one obtains
\begin{eqnarray}
v^{(0)} &=& T^{(0)} \ , \label{eq:v0}\\
v^{(1)} &=& T^{(1)}-\left[ v^{(0)}\, G_0\, v^{(0)}\right] \ , \\
v^{(2)} &=& T^{(2)}-\left[ v^{(0)}\, G_0\, v^{(0)}\, G_0\, v^{(0)}\right] \nonumber\\
&&\qquad-\left[ v^{(1)}\, G_0 \, v^{(0)}
+v^{(0)}\, G_0\, v^{(1)}\right] \ . \label{eq:v2}
\end{eqnarray}
Notice that a term like $v^{(m)} G_0 v^{(n)}$ is of order $Q^{m+n+1}$, since $G_0$ is of order $Q^{-2}$ and the implicit loop integration brings in a factor $Q^3$.
The leading-order (LO) $Q^0$ term, $v^{(0)}$, consists of two (non-derivative) contact interactions and (static) one-pion exchange (OPE) (respectively displayed in panels (a') and (b'), of Fig.~\ref{fig:fpot}),
while the next-to-leading (NLO)
$Q^1$ term, $v^{(1)}$, is easily seen to vanish~\cite{Pastore11}, since the leading non-static corrections
$T^{(1)}$ to the (static) OPE amplitude add up to zero on the energy
shell, while the remaining diagrams in $T^{(1)}$ represent iterations of $v^{(0)}$,
whose contributions are exactly canceled by $\left[ v^{(0)}\, G_0\, v^{(0)}\right]$
(complete or partial cancellations of this type persist at higher $n\ge 2$ orders).
The next-to-next-to-leading (N2LO) $Q^2$ term, which follows from Eq.~(\ref{eq:v2}),
contains contact (involving two gradients of the nucleon fields)
interactions, two-pion-exchange (TPE), loop corrections to LO contact interactions, and loop corrections to OPE potential (respectively displayed in panels (c'), (d')-(f'), (g') and (h'), and (i'), of Fig.~\ref{fig:fpot}). 
However, the procedure outlined above does not specify the potential uniquely, being affected by well known off energy-shell ambiguities. Indeed, at N2LO there is also a recoil correction
to the OPE, which we write as~\cite{Friar77}
\begin{equation}
\label{eq:3.15}
v^{(2)}_\pi(\nu)=v^{(0)}_\pi({\bf k})\, \frac{(1-\nu)\left[(E_1^\prime-E_1)^2+(E_2^\prime-E_2)^2\right]
	-2\,\nu\, (E_1^\prime-E_1)(E_2^\prime-E_2)}{2\,\omega_k^2} \ ,
\end{equation}
where $v^{(0)}_\pi({\bf k})$ is the leading order OPE potential, defined as
\begin{equation}
v^{(0)}_\pi({\bf k})=-\frac{g^2_A}{4\, f_\pi^2}\, {\bm \tau}_1\cdot{\bm \tau}_2
\,\, {\bm \sigma}_1\cdot{\bf k} \,\,{\bm \sigma}_2 \cdot{\bf k}\,\frac{1}{\omega_k^2} \ ,
\end{equation}
$E_i$ (${\bf p}_i$) and $E_i^\prime$ (${\bf p}^\prime_i$) are
the initial and final energies (momenta) of nucleon $i$, and
${\bf k}={\bf p}_1-{\bf p}^\prime_1$.
There is an infinite class of corrections $v^{(2)}_\pi(\nu)$, labeled by
the parameter $\nu$, which, while equivalent on the energy shell ($E_1^\prime+E_2^\prime=E_1+E_2$)
and hence independent of $\nu$, are different off the energy shell.
Friar~\cite{Friar77} has in fact shown that these different off-the-energy-shell
extrapolations $v^{(2)}_\pi(\nu)$ are unitarily equivalent, and thus do not affect physical observables. The off-shell ambiguity propagates to the next-order $v^{(3)}$, but the unitary equivalence persists also at this order, i.e., at the two-pion exchange level  \cite{Pastore11}.

The inclusion (in first order) of electroweak interactions in the perturbative
expansion of Eq.~(\ref{eq:pt}) is in principle straightforward.
The weak transition operator can be expanded as~\cite{Pastore11,Baroni16}:
\begin{equation}
T_5=T_5^{(n)}+T_{5}^{(n+1)}+T_{5}^{(n+2)} +\dots \ ,
\end{equation}
where $T_5^{(m)}$ is of order $Q^m$ and $n=-3$ in this case.
The nuclear weak axial charge, $\rho_{5,a}$,
and current, ${\bf j}_{5,a}$, operators follow
from $v_5= A^0_a\, \rho_{5,a}-{\bf A}_a\cdot {\bf j}_{5,a}$, where
$A_a^\mu=(A_a^0,{\bf A}_a)$ is the weak axial
field, and it is assumed that $v_5$ has a similar expansion
as $T_5$.  The requirement that, in the context of the LS equation,
$v_5$ matches $T_5$ order by order in the power counting
implies relations for $v^{(n)}_5=A^0_a\, \rho_{5,a}^{(n)}-{\bf A}_a\cdot {\bf j}_{5,a}^{(n)}$,
which can be found in Refs.~\cite{Pastore11,Baroni16}, similar to those derived above
for $v^{(n)}$, the strong-interaction potential.  
The lowest order terms that contribute to the axial current
operators have $n=-3$, while $n=-2$ for the axial charge. This implies that the off-shell ambiguity affects the axial current already at N3LO and the axial charge at N4LO. In the case of the electromagnetic operators the same was true with inverted roles of the charge and current \cite{Pastore11}. There it was  shown that  different choices for the $\nu$ parameter for both the potential and the electromagnetic charge operator were unitarily equivalent. We expect the same to occur for the axial current, although this has not been verified explicitly. The specific form of the axial current we use corresponds to the choice $\nu=0$  for $v^{(2)}_\pi(\nu)$ and $v^{(3)}_{2\pi}(\nu)$, specifically
Eq.~(\ref{eq:3.15}) above and Eq.~(19) of Ref.~\cite{Pastore11}.
The remaining non-static corrections in the potential $v^{(3)}$ are as given
in Eqs.~(B8), (B10), and (B12) of that work.

We notice that at N4LO there are several one loop diagrams that contribute to the nuclear axial current.
Diagrams (k), (l), (p), (q), and (r) of Fig.~\ref{fig:f1} are irreducible and in Ref.~\cite{Baroni16} they were shown to give the same contribution both in TOPT and HBPT. The remaining topologies contain reducible diagrams and require the subtraction of the iterations generated by the LS equation \cite{Pastore11, Piarulli13, Baroni16}. 
The partially conserved axial current (PCAC) relation implies the conservation of the weak axial current in the chiral limit
 ${\bf q}\cdot{\bf j}_{5,a}=\left[\, H\, ,\, \rho_{5,a}\,\right]$
 with the two-nucleon Hamiltonian given by
 $H=T^{(-1)}+v^{(0)}+v^{(2)}+\dots\,\,$ and
 where the (two-nucleon) kinetic energy $T^{(-1)}$
 is counted as $Q^{-1}$. This requirement,  order
 by order in the power counting, translates into a set of non-trivial relations
 between the ${\bf j}_{5,a}^{(n)}$ and the $T^{(-1)}$, $v^{(n)}$, and
 $\rho_{5,a}^{(n)}$ (note that commutators implicitly bring in factors
 of $Q^{3}$), see Eqs.~(7.9)--(7.12) of Ref.~\cite{Baroni16}.
 These relations couple contributions of different orders in the power counting
 of the operators, and can only be satisfied up to a truncation of the low-energy expansion. In Ref.~\cite{Baroni16} it has been shown that the axial current, up to order $Q$, is conserved in the chiral limit. In particular we note that the sum of the loop corrections at order $Q$ displayed in Fig.~\ref{fig:f1}, when contracted with the three momentum ${\bf q}$ of the external axial field, is equal to the following commutator 
 \begin{equation}
 \left[\,v_\pi^{(0)},\rho_{5,a}^{(-1)}\right]\ ,
\end{equation}
 where $v_\pi^{(0)}$ is the OPE potential, panel (b') of Fig.~\ref{fig:fpot}, and $\rho_{5,a}^{(-1)}$ is the LO two-body axial charge.
 Finally we note that the verification of PCAC, for nonvanishing pion mass, should come out as a natural consquence of the fact that we used chiral Lagrangians without making any approximations (besides neglecting some $1/m$ corrections at order $Q$, for further details we defer to Sec IV.B of Ref.~\cite{Baroni16}). However an explicit verification of PCAC for tree level diagrams as well as loop corrections at order $Q$ of Fig.~\ref{fig:f1} has not yet been performed.

\section{Nuclear axial currents in $\chi$EFT}
\label{sec:sec2}
In this section we report the expressions for the nuclear axial current
in the limit of vanishing external field momentum (denoted as ${\bf q}$)~\cite{Baroni16}.
Of course, pion-pole contributions in Fig.~\ref{fig:f1} vanish in this limit.
The expressions at LO and N2LO read
\begin{eqnarray}
{\bf j}^{\rm LO}_\pm&=&-g_A \, \tau_{1,\pm}\, {\bm\sigma}_1+ \left(1\rightleftharpoons 2\right) \ , 
\label{eq:alo}\\
{\bf j}_\pm^{\rm N2LO} &=&\frac{g_A}{2\, m^2}\,\tau_{1,\pm} 
\left( K_1^2\, {\bm \sigma}_1- {\bf K}_1\,\,{\bm \sigma}_1\cdot{\bf K}_1 \right) + \left(1\rightleftharpoons 2\right)\ ,
\label{eq:anlo}
\end{eqnarray}
while those at N3LO are
separated into one-pion exchange (OPE) and contact (CT) terms corresponding
respectively to panels (e) and (g) of Fig.~\ref{fig:f1},
\begin{eqnarray}
\label{eq:an2lo}
{\bf j}_\pm^{\rm N3LO}({\rm OPE};{\bf k})&=&  \frac{g_A}{2\, f_\pi^2} \left\{ 
4\, c_3 \, \tau_{2,\pm}\, {\bf k} +\left({\bm \tau}_1\times{\bm \tau}_2\right)_\pm
\left[\left( c_4+\frac{1}{4\, m}\right){\bm \sigma}_1\times{\bf k}
-\frac{i}{2\, m} {\bf K}_1 \right]\right\} \nonumber\\
&&\times  {\bm\sigma}_2\cdot{\bf k}\, \frac{1}{\omega_k^2} + \left(1\rightleftharpoons 2\right)\ , \\
{\bf j}_\pm^{\rm N3LO}({\rm CT};{\bf k})&=&z_0\, \left({\bm \tau}_1\times{\bm \tau}_2\right)_\pm\,
 {\bm \sigma}_1\times{\bm \sigma}_2\ . 
\label{eq:act} 
\end{eqnarray} 
The LECs $c_3$ and $c_4$ in the OPE current effectively include
the contributions associated with $\Delta$-isobar excitations
($\Delta$ degrees of freedom are integrated out in the $\chi$EFT formulation adopted here) as
well as short-range contributions involving vector meson exchanges,
such as axial $\rho$-$\pi$ transition mechanisms~\cite{Park03}.

Lastly, the expressions at N4LO are
separated into terms originating from OPE, panel (s), and multi-pion exchange (MPE),
panels (i), (k), (m), and (p),
\begin{eqnarray}
\label{eq:an3lo1}
{\bf j}_\pm^{\rm N4LO}({\rm OPE};{\bf k})&=& \frac{g^5_A\, m_\pi}{256\,\pi\,f_\pi^4} \left[\,
18\, \tau_{2,\pm}\, {\bf k} - \left({\bm \tau}_1\times{\bm \tau}_2\right)_\pm
{\bm \sigma}_1\times{\bf k} \,\right] {\bm\sigma}_2\cdot{\bf k}\, \frac{1}{\omega_k^2}
+ \left(1\rightleftharpoons 2\right)\ , \\
{\bf j}_\pm^{\rm N4LO}({\rm MPE};{\bf k})\!\!&=&\!\!\frac{g_A^3}{32 \, \pi f_\pi^4}\, 
\tau_{2,\pm}\left[ W_1(k)\, {\bm \sigma}_1 +W_2(k)\, {\bf k}\,\,{\bm \sigma}_1\cdot{\bf k}
+Z_1(k)\left( 2\, {\bf k}\, \,
{\bm \sigma}_2\cdot{\bf k} \frac{1}{\omega_k^2}
 - {\bm \sigma}_2\right) \right] \nonumber\\
&&+\frac{g_A^5}{32\, \pi f_\pi^4}\, \tau_{1,\pm }\, W_3(k) \left({\bm \sigma}_2\times{\bf k}\right)
\times{\bf k} -\frac{g_A^3}{32 \, \pi f_\pi^4}\left({\bm \tau}_1\times{\bm \tau}_2\right)_\pm
Z_3(k)\, {\bm \sigma}_1\times{\bf k} \nonumber\\
&&\times {\bm \sigma}_2\cdot{\bf k}\, \frac{1}{\omega_k^2} 
+ \left(1\rightleftharpoons 2\right)\ ,
\label{eq:ampe}
\end{eqnarray}
where the loop functions are given by
\begin{eqnarray}
\label{eq:ew1}
W_1(k)&=&\int_0^1 dz\left[ \left(1-5\,g_A^2\right)M(k,z)
-\frac{g_A^2 \,k^2}{2}\left[ \frac{9\, z\,\overline{z}-1}{M(k,z)}
-\frac{k^2(z\, \overline{z})^2}{M(k,z)^3}\right]\right]\ , \\
W_2(k)&=&\int_0^1 dz\left[-\frac{g_A^2\,(z\,\overline{z})^2\, k^2}{2\, M(k,z)^3}
+\frac{z\, \overline{z}\left(7\,g_A^2+2\right)-g_A^2}{2M(k,z)}\right]\ , \\
W_3(k)&=&-\frac{1}{2}\int_0^1 dz\left[ \frac{k^2\,(z-\overline{z}\,)^2}{12\, M(k,z)^3}
+\frac{1}{M(k,z)}\right]\ ,\\
Z_1(k)&=&\int_0^1 dz \left[ \frac{z\, \overline{z}\, k^2}{M(k,z)}+3\, M(k,z)\right] \ , \\
Z_3(k)&=& \int_0^1 dz \, M(k,z)\ ,
\end{eqnarray}
and
\begin{equation}
M(k,z)=\sqrt{z\overline{z}\,k^2+m_\pi^2} \ ,\qquad \overline{z}=1-z \ .
\end{equation}
In the equations above, $g_A$ and $f_\pi$ are the nucleon axial coupling
constant and pion decay amplitude, $m$ and $m_\pi$ are the nucleon
and pion mass, $\omega_k=\sqrt{k^2+m_\pi^2}$ is the pion energy, and
$c_3$, $c_4$, and $z_0$ are LECs, $c_3$ and $c_4$
entering the ${\cal L}^{(2)}_{\pi N}$ Lagrangian and
$z_0$ multiplying the contact axial current (these LECs are discussed in Sec.~\ref{sec:res}).
The nucleon spin and isospin operators are denoted by ${\bm \sigma}$
and ${\bm \tau}$, respectively, and the following charge-raising
($+$) and charge-lowering ($-$) combinations have been defined:
\begin{equation}
\tau_{i,\pm}=(\tau_{i,x}\pm i\, \tau_{i,y})/2 \ , \qquad ({\bm \tau}_1\times{\bm \tau}_2)_\pm=
({\bm \tau}_1\times{\bm \tau}_2)_x \pm i\, ({\bm \tau}_1\times{\bm \tau}_2)_y \ .
\end{equation}
The momenta ${\bf k}_i$ and ${\bf K}_i$ are
\begin{equation}
{\bf k}_i= {\bf p}_i^\prime-{\bf p}_i\ , \qquad {\bf K}_i =\left({\bf p}_i^\prime+{\bf p}_i\right)/2\ ,
\end{equation}
where ${\bf p}_i$ (${\bf p}_i^\prime$) is the nucleon initial (final) momentum and,
in the limit of vanishing external field momentum, ${\bf k}_1$
and ${\bf k}_2$ are related via
\begin{equation}
{\bf k}_1={\bf k}=-{\bf k}_2 \ .
\end{equation}

In Ref.~\cite{Baroni16} diagrams (w) and (x) of Fig.~\ref{fig:f1} were inadvertently omitted,
only diagrams (u) and (v) were considered.  We have evaluated them here,
and obtained for the combined contribution of (u) and (w) the N4LO contact
current
\begin{equation}
{\rm diagrams\,\, (u)+(w)}=-\frac{g_A^3\, m_\pi}{16\,\pi\, f_\pi^2} \,C_T \, 
\big[4\,(\tau_{1,\pm}-\tau_{2,\pm})\, {\bm \sigma}_2\, +
({\bm \tau}_1\times{\bm \tau}_2)_\pm
\left({\bm \sigma}_1\times{\bm \sigma}_2\right)\big] + \left(1\rightleftharpoons 2\right)\ ,
\label{eq:ct1}
 \end{equation} 
where $C_T$ (in standard notation) is one of the two LECs in the
four-nucleon contact interaction at LO.  The pion-pole contribution from diagrams
(v)+(x) follows as
\begin{equation}
{\rm diagrams\,\, (v)+(x)}=-\frac{{\bf q}}{q^2+m_\pi^2}\, {\bf q}\cdot \left[{\rm diagrams\,\, (u)+(w)}\right] \ .
\label{eq:ct2}
\end{equation}
However, use of Fierz identities shows that the contact current in Eq.~(\ref{eq:ct1}) vanishes
identically~\cite{Baroni16}.
\begin{figure}[bth]
\includegraphics[width=6cm]{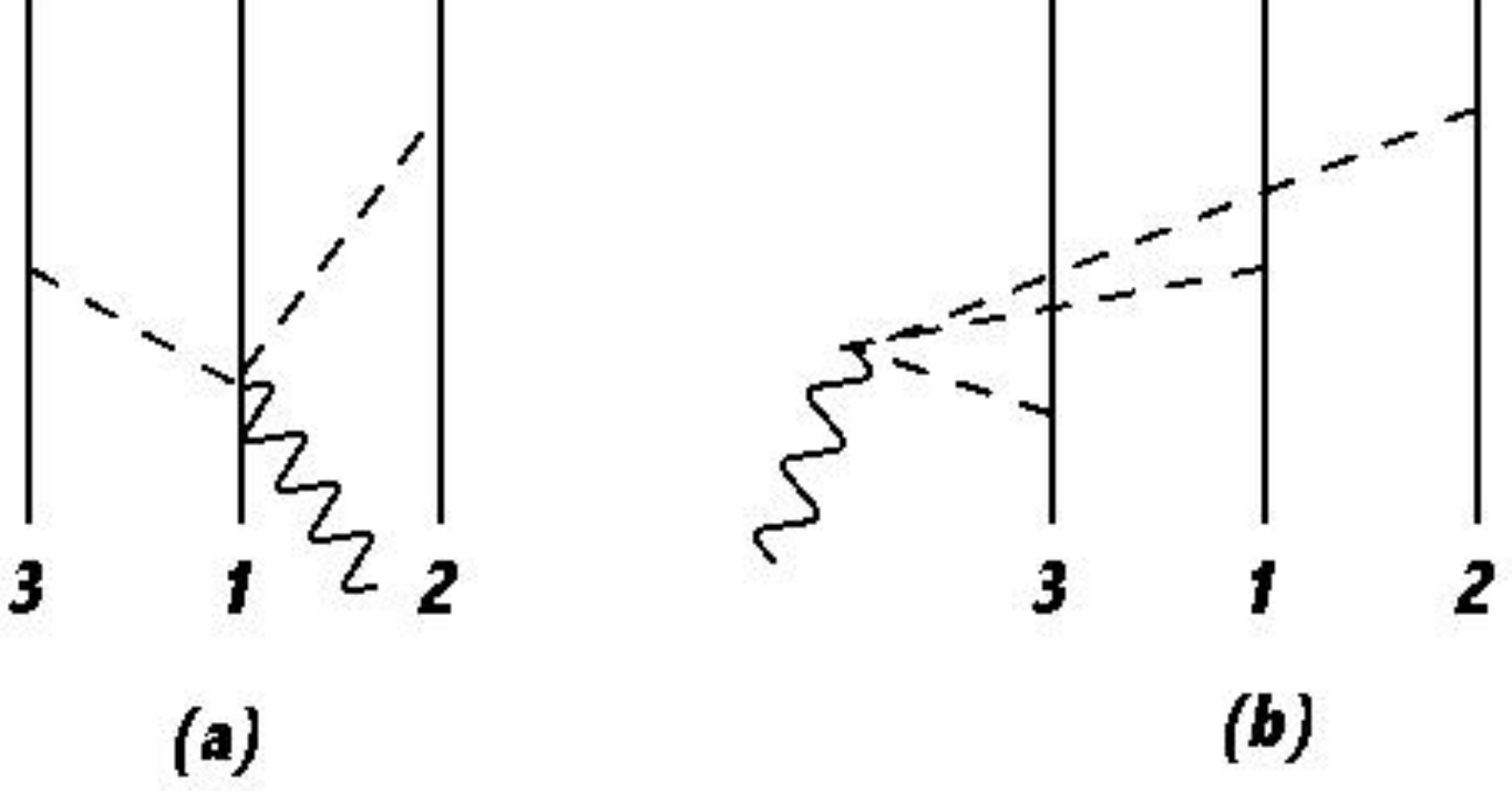}
\caption{Diagrams illustrating the three-body axial current at N4LO (i.e., order $Q^{-2}$ in
a three-nucleon system).  Nucleons, pions, and axial fields are denoted
by solid, dashed, and wavy lines, respectively.  Only a single time ordering
is shown and pion-pole contributions are ignored.}
\label{fig:f2}
\end{figure}

In a three-nucleon system the two-body loop corrections
to the axial current enter at order $Q^{-2}$, owing to the presence
of a momentum-conserving $\delta$-function $\delta({\bf p}_3^\prime-{\bf p}_3)$.
These loop corrections turn out to be of the same order as
the three-body axial current, illustrated in Fig.~\ref{fig:f2}
and first derived in Ref.~\cite{Park03},
\begin{eqnarray}
{\bf j}^{\rm N4LO}_\pm({\rm 3B};{\bf k}_2,{\bf k}_3)&=&-\sum_{\rm cyc}\frac{g_A^3}{8\,f_\pi^4}
\left(2\, \tau_{1,\pm}\,\,{\bm \tau}_2\cdot{\bm \tau}_3
-\tau_{2,\pm}\, {\bm \tau}_3\cdot{\bm \tau}_1
-\tau_{3,\pm}\, {\bm \tau}_1\cdot{\bm \tau}_2\right)\nonumber\\
&&\times\left({\bm \sigma}_1-\frac{4}{3}\frac{{\bm \sigma}_1\cdot{\bf k}_1\,{\bf k}_1}{\omega_1^2}\right)
\frac{{\bm \sigma}_2\cdot{\bf k}_2}{\omega_2^2}\frac{{\bm \sigma}_3\cdot{\bf k}_3}{\omega_3^2}
\, ,\label{eq:3b}
\end{eqnarray}
where the sum is over the cyclic permutations of the three nucleons, and in the ${\bf q}=0$ limit
${\bf k}_1=-\left({\bf k}_2+{\bf k}_3\right)$.

Configuration-space expressions for these two- and three-body operators (denoted generically
as 2B and 3B, respectively) follow from
\begin{eqnarray}
{\bf j}_\pm({\rm 2B})&=&\int \frac{d{\bf k}}{(2\pi)^3}\, {\rm e}^{i {\bf k} \cdot {\bf r}_{12} }\,\, 
C_\Lambda(k)\,\, {\bf j}({\rm 2B};{\bf k}) \ , \\
{\bf j}_\pm({\rm 3B})&=&\int \frac{d{\bf k}_2}
{(2\pi)^3}\, \frac{d{\bf k}_3}{(2\pi)^3}\,{\rm e}^{-i {\bf k}_2 \cdot {\bf r}_{12} }\,
{\rm e}^{-i {\bf k}_3 \cdot {\bf r}_{13} }\,\, 
C_\Lambda(k_2)\,\, C_\Lambda(k_3)\,\,{\bf j}({\rm 3B};{\bf k}_2,{\bf k}_3) \ ,
\end{eqnarray}
where the relative positions are defined as ${\bf r}_{ij}={\bf r}_i-{\bf r}_j$, and
$C_\Lambda(k)$ is the momentum cutoff, which we take as
\begin{equation}
C_\Lambda(k)={\rm e}^{-(k/\Lambda)^4} \ .
\label{eq:ctff}
\end{equation}
This cutoff does not modify the power counting of the various terms, as it
is easily seen by expanding in powers of $k/\Lambda$.  In particular, the conservation of
the vector current and axial current (in the chiral limit) is preserved up to the order
considered in the present work.

Lastly, terms proportional to ${\bf K}_j$ in the N2LO and N3LO currents are obtained by
replacing ${\bf K}_j$ with $-i\, {\bm \nabla}_{\!j}$ in configuration space
(the momentum operator), and need to be symmetrized accordingly to preserve
hermiticity.  Explicit expressions for these Fourier transforms are listed in
Appendix~\ref{app:a1}.
\section{Gamow-Teller matrix element in tritium $\beta$-decay}
\label{sec:res}
The Gamow-Teller (GT) matrix element is obtained from the tritium half-life via
(see~\cite{Schiavilla98} and references therein)
\begin{eqnarray}
(1+\delta_R)\, 	t\, f_V&=&\frac{K/G_V^2}
{\langle{\bf F}\rangle^2+f_A/f_V\,g_A^2\,\langle{\bf GT}\rangle^2}\ ,
\label{eq:gtf}
\end{eqnarray}
where $g_{A}=1.2723$ is the current experimental value~\cite{g_A} for the nucleon axial
coupling constant, $\delta_R=1.9\%$ is the outer radiative correction~\cite{d_R}, $t$ is the
half-life of $^3{\rm H}$, and $f_V$ and $f_A$ are Fermi functions reported in Ref.~\cite{Simpson87}
to have the values $2.8355\times 10^{-6}$ and $2.8505\times 10^{-6}$, respectively.  The
experimental value used for $K/G_V^2$ is $(6144.5\pm 1.9)$~s as obtained from Ref.~\cite{Hardy15},
and that used for $(1+\delta_R)	\,t\, f_V$ is $(1134.6\pm3.1)$~s as reported in Ref.~\cite{Simpson87}.  Finally,
$\langle {\bf F}\rangle$ and $\langle{\bf GT}\rangle$ denote the reduced matrix element of the
Fermi (F) and GT operators. The GT operator is the axial current constructed
in Sec.~\ref{sec:sec2}.  The F operator is the vector charge and, while it too includes
one- and two-body terms derived in Ref.~\cite{Pastore11}, the latter vanish in the limit of vanishing external
field momentum, and only the one-body term at LO contributes in this limit.

The F and GT matrix elements are calculated with $^3{\rm H}$ and $^3{\rm He}$ wave functions
obtained with the hyperspherical-harmonics (HH) expansion method (see review~\cite{Kievsky08})
from two- and three-nucleon potentials derived from either $\chi$EFT or the conventional approach.
The combination of chiral potentials is denoted as N3LO/N2LO(500) [N3LO/N2LO(600)] corresponding
to cutoff $\Lambda=500$ MeV ($\Lambda=600$ MeV), and consists of two-nucleon potentials at N3LO
from Refs.~\cite{Entem03,Machleidt11} and three-nucleon potentials at N2LO
from Refs.~\cite{Epelbaum02,Nav07}.~\footnote{Note that for consistency with the convention adopted
in Fig.~\ref{fig:f1}, it would be more appropriate to label these two- and three-nucleon
potentials, respectively, as  N4LO and N3LO.  However, this is not the standard notation
used in the literature.}
The combination of conventional potentials is denoted as AV18/UIX and consists of the Argonne
$v_{18}$ (AV18) two-nucleon potential~\cite{Wiringa95}  and Urbana-IX (UIX) three-nucleon
potential~\cite{Pudliner95}.  In all cases we obtain $\langle {\bf F}\rangle=0.9998$.
From this value we extract via Eq.~(\ref{eq:gtf}) the experimental GT matrix element
as
\begin{equation}
\label{eq:gtexp}
{\rm GT}_{\rm EXP}=\langle {\bf GT}\rangle_{\rm EXP}/\sqrt{3}=0.9511\, \pm\, 0.0013 \ .
\end{equation}

Contributions to the GT matrix element corresponding to the LO, N2LO, N3LO, N4LO,
and N4LO(3Ba) axial operators are reported in Table~\ref{tb:tb1}, where
the LEC $z_0$ in the N3LO(CT) operator is taken as $z_0=1$ in units of ${\rm GeV}^{-3}$.
The LECs $c_3$ and $c_4$ in the N3LO(OPE) operators are
constrained by fits to $\pi N$ scattering data, and two different
sets of values (listed in the table caption) have been used in the present study, one
from Refs.~\cite{Entem03,Machleidt11} and the other from a recent analysis of these data based
on Roy-Steiner equations~\cite{Hoferichter15}, specifically the values corresponding
to the column labeled N3LO in Table II of that work.  The first set of $c_3$ and $c_4$
values (from Refs.~\cite{Entem03,Machleidt11}) enters the chiral two- and three-nucleon
potentials, used here to generate the $^3$H and $^3$He wave functions.  Clearly,
use of the second set from Ref.~\cite{Hoferichter15} in the N3LO(OPE) axial
current is not consistent with these potentials; results for the GT matrix
element are provided in that case only to give an estimate of the their sensitivity
to the $c_3$ and $c_4$ values. As per the additional LECs $(c_D,c_E)$
in the three-nucleon potential, these have been obtained by the fitting procedure
described below.  In particular, we note that the LEC $z_0$ in the N3LO(CT) operator
is related to $c_D$ via Eq.~(\ref{eq:cdv}).

In the N4LO(3Ba) current we have only considered
the term ${\bf j}^{\rm N4LO}_\pm({\rm 3B},{\rm a})$
of Eq.~(\ref{eq:a11}) and neglected the term ${\bf j}^{\rm N4LO}_\pm({\rm 3B},{\rm b})$
of Eq.~(\ref{eq:a17}) for reasons explained in Appendix~\ref{app:a1}.  The GT (and F) matrix
elements are computed exactly, without approximation, with quantum
Monte Carlo methods.  The spin-isospin algebra is carried out with techniques
similar to those developed in Ref.~\cite{Schiavilla89} for the electromagnetic
current operator.  The results reported in the tables below are based
on random walks consisting of $10^6$ configurations.  Statistical errors are
not listed, but are typically at the few parts in $10^3$, except in the special
case of the N3LO(OPE) results, for which they are at the few \% level (see below). 

In Table~\ref{tb:tb1} we report the results for the N3LO/N2LO(500)
and N3LO/N2LO(600) models, and in parentheses those for the AV18/UIX model.
The LO and N2LO axial operators do not need to be regularized,
and hence the corresponding contributions for the AV18/UIX are the same for $\Lambda=500$
MeV and 600 MeV.  However, the N3LO/N2LO contributions change (rather significantly
at N2LO) as $\Lambda$ varies in this range due to the intrinsic cutoff dependence
of the potentials.  In the N3LO axial current of Eq.~(\ref{eq:an2lo}) the terms proportional
to $c_3$ and $c_4$ have opposite signs and tend to cancel each other.  This cancellation depends
crucially on the values of the LECs and Hamiltonian model.  In particular, when
$c_3$ and $c_4$ are taken from Refs.~\cite{Entem03,Machleidt11}, the sum of their contributions
for the N3LO/N2LO model is (in magnitude) comparable to the contribution from
the non-local terms proportional to ${\bf K}_i$ in Eq.~(\ref{eq:an2lo}).

The contributions from loop corrections, row labeled N4LO(MPE), are relatively
large and comparable to those at N3LO(OPE).  As a matter of fact, when the
values for the $c_3$ and $c_4$ LECs are from Refs.~\cite{Entem03,Machleidt11}, the
N3LO(OPE) contributions are an order of magnitude smaller than the N4LO(MPE)
in the case of the chiral potentials.  The origin of this large
contribution can be traced back to the term proportional to  the loop
function $W_1(k)$ in Eq.~(\ref{eq:ampe}), specifically to the term
with the factor $(1-5\, g_A^2)$ in Eq.~(\ref{eq:ew1}).  It originates from
box diagrams, panel (m) of Fig.~\ref{fig:f1} (see Ref.~\cite{Baroni16}).  All the
N4LO corrections have opposite signs relative to the LO and N3LO(OPE).
\begin{table*}[bth]
\caption{Contributions to the GT matrix element of tritium $\beta$-decay corresponding to
the Hamiltonian model N3LO/N2LO (AV18/UIX) and cutoffs $\Lambda=500$ MeV and
600 MeV in the chiral potentials and weak axial current operators.  The acronyms LO, N2LO,
N3LO(OPE), N3LO(CT), N4LO(OPE), N4LO(MPE), and N4LO(3Ba) refer, respectively,
to the axial operators given in Eq.~(\ref{eq:alo}), Eq.~(\ref{eq:anlo}),
Eq.~(\ref{eq:an2lo}), Eq.~(\ref{eq:act}), Eq.~(\ref{eq:an3lo1}),
Eq.~(\ref{eq:ampe}), and Eq.~(\ref{eq:a11}).  In the N3LO(OPE) operator
the LECs $c_3$ and $c_4$ have the values
$c_3=-3.20\,{\rm GeV}^{-1}$ and $c_4=5.40\,{\rm GeV}^{-1}$ from Refs.~\cite{Entem03,Machleidt11}, while
in the N3LO$^\star$(OPE) operator they  are taken as $c_3=-5.61\,{\rm GeV}^{-1}$ and
$c_4=4.26\,{\rm GeV}^{-1}$ from Ref.~\cite{Hoferichter15}. The LEC $z_0$ in N3LO(CT) is taken to have the value $z_0=1$ in units of ${\rm GeV}^{-3}$.
The LECs $(c_D,c_E)$ in the three-nucleon chiral potential have the values 
$(-1.847,-0.548)$ for
$\Lambda\,$=$\, 500$ MeV and  $(-2.030,-1.553)$
for $\Lambda=600$ MeV. See text for further explanations.}
\begin{tabular}{c|c|c}
$\Lambda$& 500 MeV & 600 MeV \\
\hline 
\hline
LO& 0.9363(0.9224)& 0.9322 (0.9224) \\
\hline
N2LO&--0.569(--0.844)$\times 10^{-2}$& --0.457(--0.844)$\times 10^{-2}$  \\
\hline
N3LO(OPE) &0.825(1.304)$\times 10^{-2}$  & 0.043(7.517)$\times 10^{-2}$ 	\\
N3LO$^\star$(OPE) & 0.579(0.812)$\times10^{-1}$ & 0.652(1.413)$\times 10^{-1}$	\\
\hline
N3LO(CT) & --0.586(--0.721)$\times 10^{-3}$ & --0.717(--0.644)$\times 10^{-3}$ \\ 
\hline
N4LO(OPE) & --0.697(--0.964)$\times 10^{-2}$  & --0.867(--1.216)$\times 10^{-2}$ 	\\
\hline
N4LO(MPE) & --0.430(--0.565)$\times 10^{-1}$ & --0.532(--0.775)$\times 10^{-1}$\\
\hline
N4LO(3Ba)& --0.143(--0.183)$\times 10^{-2}$  & --0.153(--0.205)$\times 10^{-2}$  \\
\hline
\end{tabular}
\label{tb:tb1}
\end{table*}

Next, we discuss the determination of the value for the LEC $z_0$ required to reproduce
GT$_{\rm EXP}$ for the various Hamiltonian models we consider, by
retaining corrections in the axial current up to either N3LO or N4LO.  In order to
compare with previous determinations of this LEC~\cite{Park03,Gazit09,Marcucci12},
we define an adimensional
$\hat{z}_0$ by rescaling $z_0$ as
\begin{equation}
\hat{z}_0=\frac{2\, m\, f_\pi^2}{g_A}\, z_0 \ .
\end{equation}
This $\hat{z}_0$ is simply given by $\hat{z}_0=\hat{d}_1+2\, \hat{d}_2$
in terms of the LECs $\hat{d}_1$ and $\hat{d}_2$ introduced in Ref.~\cite{Park03}
(in~\cite{Park03} these LECs multiply contact axial currents related to each other by a Fierz
rearrangement, and are not therefore independent).  We also note the relation
\begin{equation}
\hat{d}_R=\hat{z}_0+\frac{\hat{c}_3}{3}+\frac{2\,\hat{c}_4}{3}+\frac{1}{6}\ ,\label{eq:dr}
\end{equation}
where $\hat{c}_i=m\,c_i$ are adimensional, and $\hat{d}_R$ was fixed in Ref.~\cite{Park03}
by fitting GT$_{\rm EXP}$ in a hybrid calculation based on the AV18/UIX
model and including N3LO corrections in the axial current.
\begin{table}[bth]
\caption{Adimensional values of the LECs corresponding to
the AV18/UIX Hamiltonian model and cutoffs $\Lambda=500$ MeV and
600 MeV in the chiral axial current.  The LEC $\hat{z}_0$ is
determined by reproducing GT$_{\rm EXP}$ in calculations including in this
current corrections up to either N3LO or N4LO.  The values for $\hat{z}_0$, $\hat{d}_R$, and $c_D$
are obtained using the LECs $(c_3,c_4)\,$=$\,(-3.20,5.40)$ GeV$^{-1}$ from Refs.~\cite{Entem03,Machleidt11},
those for $\hat{z}_0^\star$, $\hat{d}_R^\star$, and $c_D^\star$ 
using $(c_3,c_4)\,$=$\,(-5.61,4.26)$ GeV$^{-1}$ from Ref.~\cite{Hoferichter15},
in both the N3LO and N4LO calculations.
}
\begin{tabular}{c|c|c||c|c}					
& \multicolumn{2}{c||}{N3LO} &
\multicolumn{2}{c}{N4LO}\\
\hline
$\Lambda$&  500 & 600 & 500& 600 \\
\hline
$\hat{z}_0$&--0.421& 0.742& --1.607& --1.048\\
\hline
$\hat{d}_R$& 2.122& 3.285&$\,\,$0.936&$\,\,$ 1.495\\
\hline
$c_D$&--0.571&1.007&--2.180 &--1.421\\
\hline\hline
$\hat{z}_0^{\star}$& 0.769&2.038&--0.417&0.235\\
\hline
$\hat{d}_R^{\star}$& 1.850 &3.115 &0.660 &1.311\\
\hline
$c_D^{\star}$&1.043 &2.764&--0.566&0.318\\	
\hline
\end{tabular}
\label{tb:tb2}
\end{table} 
Lastly, the LEC $c_D$ in the three-nucleon potential at N2LO
is related to $\hat{z}_0$ via~\cite{Park03,Gazit09,Marcucci12}
\begin{equation}
\label{eq:cdv}
c_D=\frac{g_A\, \Lambda_\chi}{m}\hat{z}_0\ ,
\end{equation}
where $\Lambda_{\chi}$ is taken as 1 GeV here, while in Refs.~\cite{Gazit09,Marcucci12}
$\Lambda_{\chi}=0.7$ GeV was adopted
($\Lambda_\chi$ is not to be confused
with the cutoff $\Lambda$ which regularizes the configuration-space
expressions of the axial operators).

\begin{figure}[t]
\vspace*{0.5cm}
\includegraphics[width=4in]{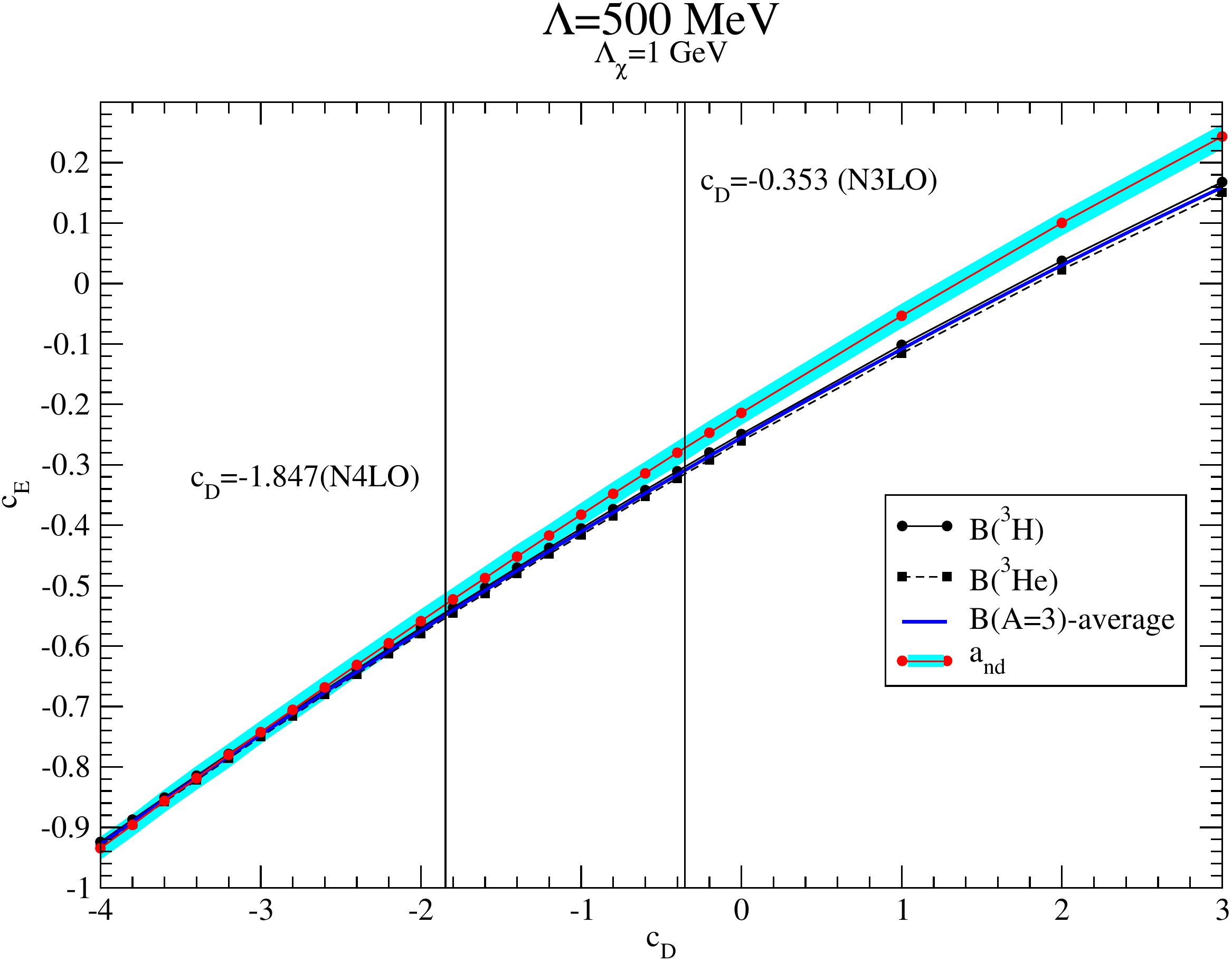}
\vspace*{0.1cm}
\caption{The $c_D$-$c_E$ trajectories fitted to reproduce the experimental
$A\,$=$\,3$ binding energies and the doublet $nd$ scattering length
using the N3LO/N2LO potential with $\Lambda\,$=$\,500$ MeV.
The values of 8.475 MeV, 7.725 MeV and $0.645\pm 0.010$ fm~\cite{Schoen03} 
are used for the 
$^3$H, $^3$He and $nd$ scattering length, respectively.
Note that the $A=3$ binding energies have been corrected for the small
contributions ($+7$ keV in $^3$H and $-7$ keV in $^3$He) due to the
$n$-$p$ mass difference~\cite{Nogga03}. The (cyan) band is due to the 
experimental uncertainty on the $nd$ scattering length. The vertical lines indicate the $c_D$
values obtained by fitting GT$_{\rm EXP}$ and retaining N4LO or only N3LO
contributions in the axial current are also displaied.}
\label{fig:be-as-fit500}
\end{figure}

\begin{figure}[t]
\vspace*{0.5cm}
\includegraphics[width=4in]{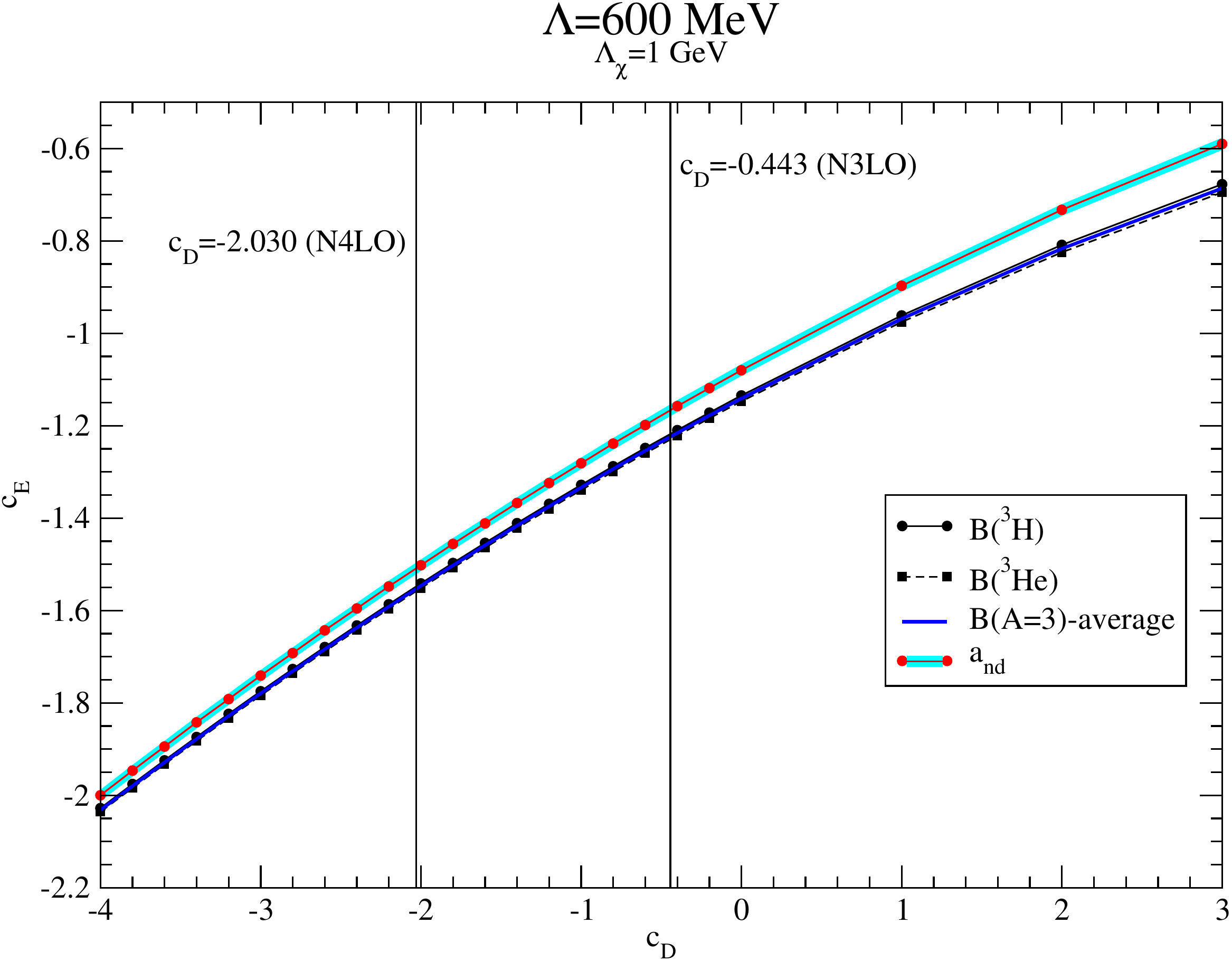}
\vspace*{0.1cm}
\caption{Same as Fig.~\ref{fig:be-as-fit500} but for $\Lambda\,$=$\, 600$ MeV.}
\label{fig:be-as-fit600}
\end{figure}

\begin{figure}[t]
\vspace*{0.5cm}
\includegraphics[width=4in]{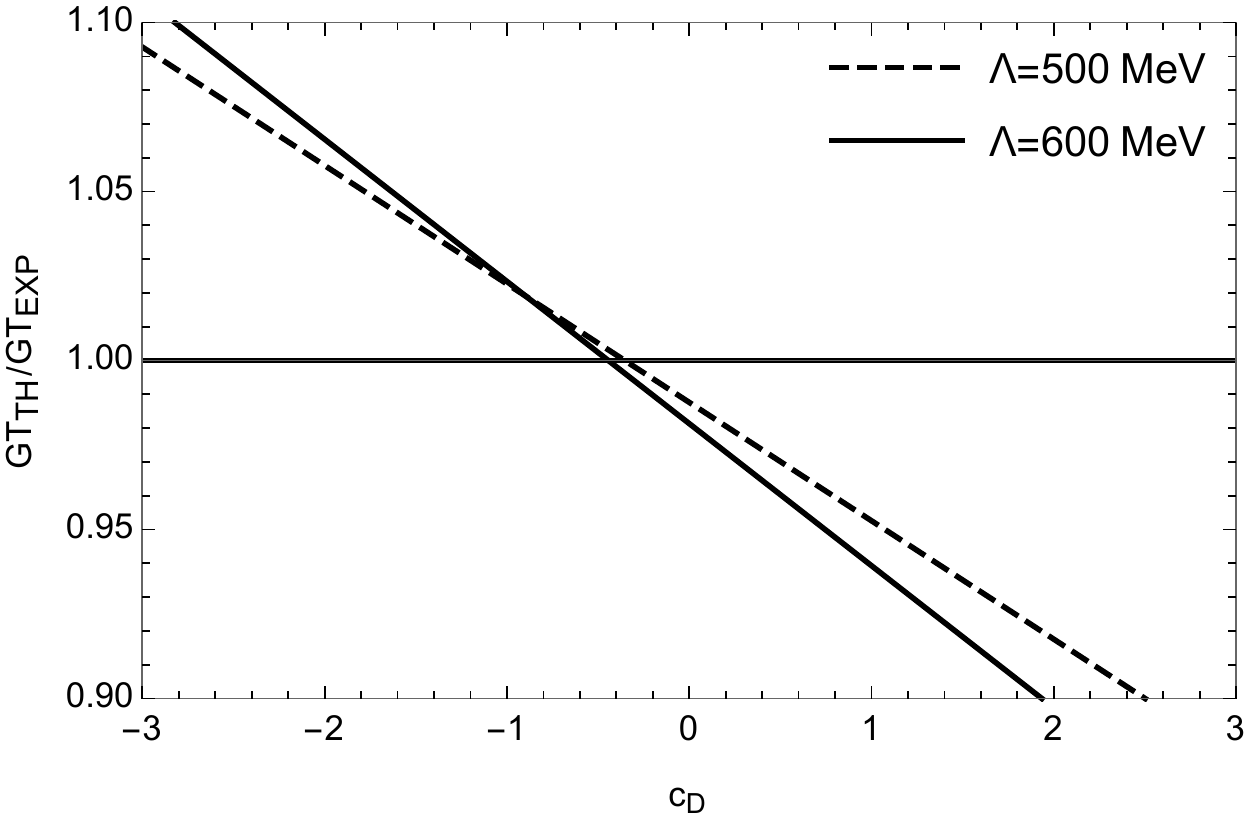}
\vspace*{0.1cm}
\caption{The ratio GT$_{\rm TH}$/GT$_{\rm EXP}$ as function of the LEC $c_D$
obtained retaining corrections up to N3LO in the nuclear axial
current. The results for both values of the cutoff $\Lambda$ are shown.}
\label{fig:gtN3LO}
\end{figure}

\begin{figure}[t]
\vspace*{0.5cm}
N4Ll\includegraphics[width=4in]{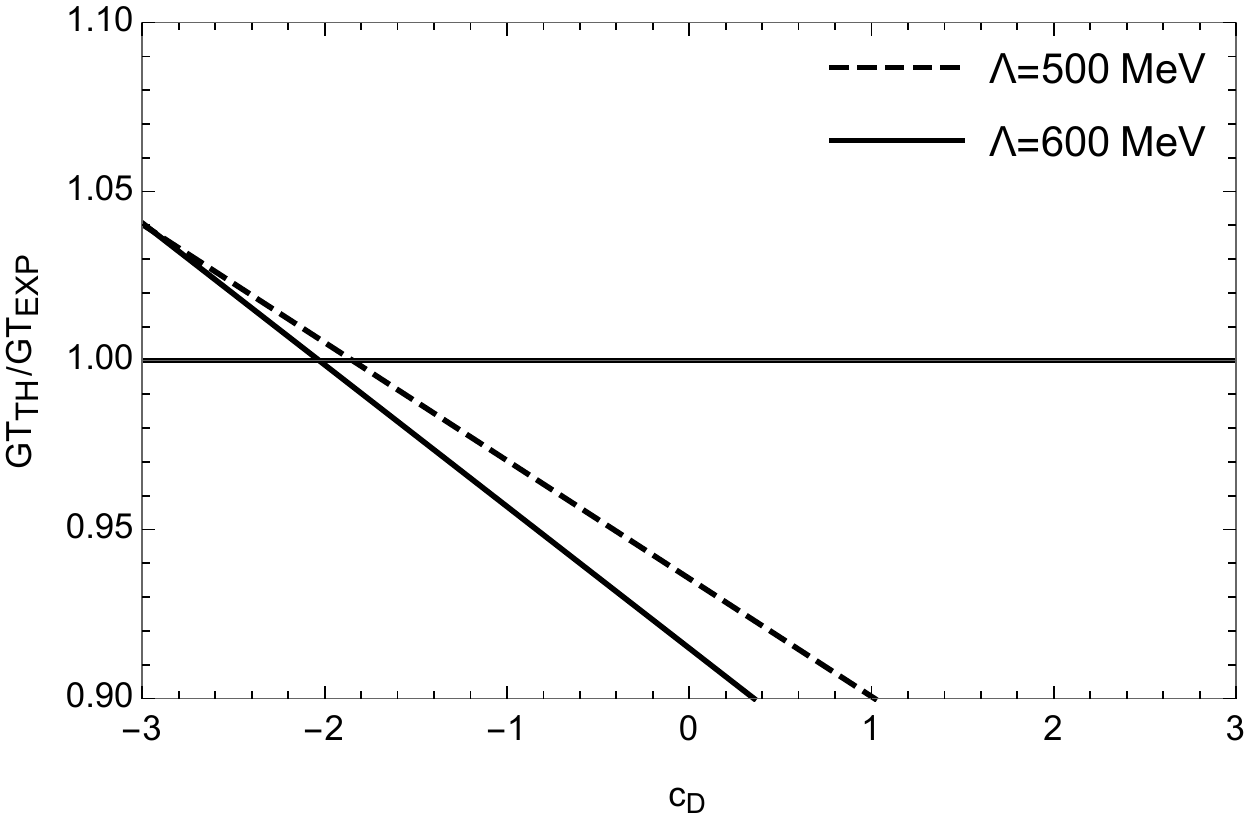}
\vspace*{0.1cm}
\caption{Same as Fig.~\ref{fig:gtN3LO} but with the corrections
in the axial current up to N4LO.}
\label{fig:gtN4LO}
\end{figure}

Values for the LECs are reported in Table~\ref{tb:tb2}
for the hybrid calculation based on the AV18/UIX Hamiltonian model,
and in Table~\ref{tb:cd-ce} for the chiral Hamiltonian model.
In Table~\ref{tb:tb2} the values for the various combinations considered above
are listed, so that they can be compared with
previous determinations~\cite{Park03,Marcucci12,Marcucci11}: they
follow simply from reproducing the central value of GT$_{\rm EXP}$
in Eq.~(\ref{eq:gtexp}).
In order to determine the values corresponding to the
chiral potentials, we proceed as in Ref.~\cite{Marcucci12}.
The $^3$H and $^3$He ground state wave functions are calculated using these potentials
for $\Lambda\,$=$\,500$ MeV and 600 MeV.  We span the range $c_D\in[-4,3]$,
and, in correspondence to each $c_D$ in this range, determine $c_E$ so
as to reproduce the binding energies of either $^3$H or $^3$He. 
The resulting trajectories are essentially indistinguishable, as shown
in Fig.~\ref{fig:be-as-fit500} for $\Lambda\,$=$\,500$ MeV and in
Fig.~\ref{fig:be-as-fit600} for $\Lambda\,$=$\, 600$ MeV, and as
already obtained in Ref.~\cite{Marcucci12}.  Then, for each set of
$(c_D,c_E)$, the triton and $^3$He wave functions are calculated
and the Gamow-Teller matrix element, denoted as GT$_{\rm TH}$, is
determined, by including in the axial current corrections up to N3LO or N4LO.
The ratio GT$_{\rm TH}$/GT$_{\rm EXP}$ for both values
of the cutoff $\Lambda$ is shown in Fig.~\ref{fig:gtN3LO} for the N3LO
case and Fig.~\ref{fig:gtN4LO} for the N4LO one.  The LECs $(c_D,c_E)$ that
reproduce GT$_{\rm EXP}$ (its central value) and the trinucleon
binding energies are given in Table~\ref{tb:cd-ce}.
The values for $c_D$ at N3LO are found to be
consistent with those listed in~\cite{Marcucci12}, after allowance is made
for the different $\Lambda_\chi$ (0.7 GeV in that work versus 1 GeV above)
and for the fact that GT$_{\rm EXP}$ as determined here is slightly smaller
than adopted in~\cite{Marcucci12}. 

\begin{table}[bth]
\caption{Values for the $(c_D,c_E)$ LECs as obtained by fitting the
$A=3$ binding energy and GT$_{\rm EXP}$ (its central value), using the N3LO/N2LO
potential models with cutoffs $\Lambda=500$ MeV and 600 MeV. The results
labelled N3LO and N4LO are obtained retaining in the nuclear axial current
up to N3LO and N4LO contributions, respectively.}
\begin{tabular}{c|c|c||c|c}					
& \multicolumn{2}{c||}{N3LO} &
\multicolumn{2}{c}{N4LO}\\
\hline
$\Lambda$&  500 & 600 & 500& 600 \\
\hline
$c_D$&--0.353&--0.443& --1.847& --2.030\\
\hline
$c_E$&--0.305&--1.224&--0.548&--1.553\\
\hline
\hline
\end{tabular}
\label{tb:cd-ce}
\end{table} 


Alternatively, we could choose a different set of three-nucleon observables
to fit these LECs. We consider here, together with the $A\,$=$\,3$ binding
energy, the $nd$ doublet scattering length $a_{nd}$, 
for which we take the experimental value $0.645\pm 0.010$ fm,
obtained in Ref.~\cite{Schoen03}.  In the range $c_D\in[-4,3]$ the resulting trajectories
are displayed in Figs.~\ref{fig:be-as-fit500} and~\ref{fig:be-as-fit600}
for $\Lambda=500$ MeV and 600 MeV, respectively.  
The experimental uncertainty in $a_{nd}$ has been taken into account, and therefore
the results of Figs.~\ref{fig:be-as-fit500} and~\ref{fig:be-as-fit600} are
presented as a band.
The trajectories originating from the $A\,$=$\, 3$ binding energies and $nd$
scattering length are quite close to each other, but do not overlap.  
In the $\Lambda = 500$ MeV case, there is a crossing point at
$(c_D,c_E)\,$=$(-2.340,-0.567)$, while for $\Lambda = 600$ MeV there is
no crossing.  In particular, using the $(c_D,c_E)$ in Table~\ref{tb:cd-ce},
we obtain $a_{nd}=0.654 (0.665)$ fm for $\Lambda\,$=$\,500$ MeV and
$a_{nd}=0.687 (0.699) $ fm for $\Lambda\,$=$\,600$ MeV,
when the N4LO (N3LO) contributions in the  axial current are retained.
The present calculations of the $nd$ scattering
wave functions ignore higher order electromagnetic interaction terms, such as those
associated with the nucleons' magnetic moments.  These terms are known to reduce
the $a_{nd}$ value of about 3 \%~\cite{Kievsky08}, 
when the AV18/UIX Hamiltonian model is used. 
Thus, the present analysis seems to indicate that the three $A\,$=$\,3$ observables
($A\,$=$\,3$ binding energies, GT$_{\rm EXP}$, and $a_{nd}$) are simultaneously
reproduced, at least for $\Lambda = 500$ MeV, 
when the nuclear axial current retains corrections up to N4LO.

\section{Conclusions}
To summarize, in the present work we have carried out a calculation
of the F and GT matrix elements in $^3$H $\beta$-decay
with the charge-changing weak current recently derived in $\chi$EFT
up to N4LO (one loop).  The trinucleon wave functions have been obtained
from accurate hyperspherical harmonics solutions of the Schr\"odinger equation
corresponding to either chiral (N3LO/N2LO) or conventional (AV18/UIX) nuclear
potentials, and the relevant matrix elements have been computed by
Monte Carlo integration methods without any approximations (statistical
errors are typically at the level of a few parts in $10^3$).

We find that the OPE contributions at N3LO proportional to $c_3$ and
$c_4$ interfere destructively and therefore depend strongly on the
values of these LECs.  As a consequence, the N4LO contributions
turn out to be comparable (in magnitude) to the N3LO ones, even
though nominally they are suppressed by a factor of $Q/\Lambda_\chi$ relative
to N3LO.  This leads to a strong variation of the LEC $z_0$ as determined
respectively at N3LO or at N4LO.  It is possible that the convergence of
the chiral series is not satisfactory for this observable and that the effective
theory should be enlarged to include explicit $\Delta$'s.  An additional caveat
is that, strictly speaking, the N4LO axial current calculations reported here should
have involved the three-nucleon interaction at N3LO, whereas only the
N2LO component has been considered in this work.  Furthermore, the
definition of the current operator is closely related to the prescription adopted
for defining the nuclear potential off the energy-shell~\cite{Baroni16}.  Whether
different prescriptions lead to the same convergence pattern is a question
that would require further investigation.

Finally, the LEC multiplying the contact axial current is related to the
LEC $c_D$ in the three-nucleon potential.  This $c_D$ and the other
LEC $c_E$ which fully characterize this (contact) potential have been
constrained by a simultaneous fit to the empirical values of the three-nucleon
binding energies and GT matrix element.  When the fit is carried out
in a calculation including the axial current at N4LO, the resulting
$c_D$ and $c_E$ also lead to a doublet $nd$ scattering length in
reasonable agreement with the experimental value for $\Lambda\,$=$\,500$ MeV.

\section*{Acknowledgments}
An email exchange with B.\ Kubis in reference to the $c_i$ LECs is
gratefully acknowledged.  This research is supported by the U.S.~Department of Energy, Office of
Nuclear Physics, under contract DE-AC05-06OR23177 (A.B.~and R.S.).
A.B. was supported by a Jefferson Science Associates Theory Fellowship.
\appendix
\section{Configuration-space expressions}
\label{app:a1}
The Fourier transforms of two-body operators are easily reduced to one-dimensional
integrals [or two-dimensional ones in the case of the N4LO(MPE) operator], which are then evaluated
by Gaussian quadrature formulae.  For example, the N3LO(OPE) current is given by
\begin{equation}
{\bf j}_\pm^{\rm N3LO}({\rm OPE})={\bf j}_\pm^{\rm N3LO}(c_3)+{\bf j}_\pm^{\rm N3LO}(c_4)
+{\bf j}_\pm^{\rm N3LO}({\rm nl}) \ ,
\end{equation}
where
\begin{eqnarray}
{\bf j}^{\rm N3LO}_\pm(c_3) &=&-  \tau_{2,\pm}\left[
\frac{F_1(z;c_3)}{z}\, {\bm \sigma}_2+
F_2(z;c_3) \,\hat{\bf z}\,\left( {\bm \sigma}_2\cdot \hat{\bf z}\right)\right]+\left( 1 \rightleftharpoons 2\right) \ ,\\
{\bf j}^{\rm N3LO}_\pm(c_4) &=&-\left({\bm \tau}_1\times{\bm \tau}_2\right)_\pm\, 
{\bm \sigma}_1\times 
\left[\frac{F_1(z;c_4)}{z}\, {\bm \sigma}_2+
F_2(z;c_4) \,\hat{\bf z}\,\left( {\bm \sigma}_2\cdot \hat{\bf z}\right)\right]+\left( 1 \rightleftharpoons 2\right)\ ,\\
{\bf j}^{\rm N3LO}_\pm({\rm nl}) &=& -\left({\bm \tau}_1\times{\bm \tau}_2\right)_\pm\, 
 \left\{\, -i\,{\bm \nabla}^z_1\, , \, F_1(z;{\rm nl}) \,\,{\bm \sigma}_2\cdot \hat{\bf z}\,\right\}+\left( 1 \rightleftharpoons 2\right)\ .
\end{eqnarray}
Here we have defined ${\bf r}={\bf r}_1-{\bf r}_2$, the adimensional variable ${\bf z}=\Lambda\, {\bf r}$,
$-i\, {\bm \nabla}_i^z$ as the adimensional momentum operator, and the radial functions
\begin{eqnarray}
F_1(z;c_3)&=&-\frac{1}{\pi^2} \,\frac{g_A\, \overline{c}_3}{\overline{f}_\pi^{\,2}}\int_0^\infty dx\,
\frac{x^3}{x^2+\overline{m}_\pi^{\,2}} \, {\rm e}^{-x^4} \, j_1(xz)\ , \\
F_2(z;c_3)&=& \frac{1}{\pi^2} \,\frac{g_A\, \overline{c}_3}{\overline{f}_\pi^{\,2}} \int_0^\infty dx\,
\frac{x^4}{x^2+\overline{m}_\pi^{\,2}} \, {\rm e}^{-x^4} \, j_2(xz) \ ,
\end{eqnarray}
where $j_n(xz)$ are spherical Bessel functions.  We have also introduced adimensional constants (denoted
with the overline) expressing them units of the cutoff $\Lambda$.  They are given by
\begin{equation}
\overline{m}_\pi=m_\pi/\Lambda\ , \qquad \overline{m}=m/\Lambda\ , \qquad \overline{f}_\pi=f_\pi/\Lambda \ , \qquad
\overline{c}_3=c_3\,\Lambda \ , \qquad \overline{c}_4=c_4\,\Lambda  \ .
\end{equation}
The functions $F_1(z;c_4)$ and $F_2(z;c_4)$, and $F_1(z;{\rm nl})$ follow
from those above by the replacement of the pre-factor as 
\begin{eqnarray}
&&\frac{1}{\pi^2} \,\frac{g_A\, \overline{c}_3}{\overline{f}_\pi^{\,2}} \longrightarrow
\frac{1}{4\,\pi^2} \,\frac{g_A}{\overline{f}_\pi^{\,2}} \left(\overline{c}_4+\frac{1}{4\, \overline{m}}\right) 
\,\, {\rm for} \,\,F_1(z;c_4)\,\, {\rm and} \,\, F_2(z;c_4) \ , \\
&&\frac{1}{\pi^2} \,\frac{g_A\, \overline{c}_3}{\overline{f}_\pi^{\,2}} \longrightarrow
\frac{1}{16\,\pi^2} \,\frac{g_A}{\overline{m}\,\overline{f}_\pi^{\,2}} \,\, {\rm for} \,\,F_1(z;{\rm nl})\ .
\end{eqnarray}
The Fourier transform of the three-body operator is more involved.  We express
it as 
\begin{equation}
{\bf j}^{\rm N4LO}_\pm({\rm 3B})= {\bf j}^{\rm N4LO}_\pm({\rm 3B},{\rm a})
+ {\bf j}^{\rm N4LO}_\pm({\rm 3B},{\rm b}) \ ,
\end{equation}
where
\begin{eqnarray}
{\bf j}^{\rm N4LO}_\pm({\rm 3B},{\rm a})&=&\sum_{\rm cyc}
\left(2\, \tau_{1,\pm}\,\,{\bm \tau}_2\cdot{\bm \tau}_3
-\tau_{2,\pm}\, {\bm \tau}_3\cdot{\bm \tau}_1
-\tau_{3,\pm}\, {\bm \tau}_1\cdot{\bm \tau}_2\right) \nonumber\\
&&\times {\bm \sigma}_1 \left({\bm \sigma}_2\cdot\hat{{\bf z}}_{12}\right)
\left({\bm \sigma}_3\cdot\hat{{\bf z}}_{13}\right) F_1(z_{12};{\rm 3B})\, F_1(z_{13};{\rm 3B}) \ ,
\label{eq:a11}
\end{eqnarray}
and the function $F_1(z;{\rm 3B})$ is obtained from $F_1(z;c_3)$ by
replacing
\begin{equation}
\frac{1}{\pi^2} \,\frac{g_A\, \overline{c}_3}{\overline{f}_\pi^{\,2}} \longrightarrow
\frac{1}{4\sqrt{2}\, \pi^2} \,\frac{g^{3/2}_A}{\overline{f}_\pi^{\,2}} \ .
\end{equation}

In order to reduce the Fourier transform of the b term in the N4LO(3B) current
to a two-dimensional parametric integral,
we first regularize it as
\begin{eqnarray}
&&{\bf j}^{\rm N4LO}_\pm({\bf k}_2,{\bf k}_3;{\rm 3B},{\rm b})=\sum_{\rm cyc} \frac{g_A^3}{6\,f_\pi^4} ({\rm isospin})\,\,
{\bm \sigma}_3\cdot{\bm \nabla}_3\,\, {\bm \sigma}_2\cdot{\bm \nabla}_2\,\,
{\bm \sigma}_1\cdot{\bm \nabla}_1\,\, {\bm \nabla}_1 \,\,I\ ,\\
&&I=\int \frac{d{\bf k}_2}{(2\pi)^3}\, \frac{d{\bf k}_3}{(2\pi)^3}\,
C_\Lambda(\mid\! {\bf k}_2+{\bf k}_3\!\mid)\, {\rm e}^{-i\left({\bf k}_2\cdot{\bf r}_{12}
+{\bf k}_3\cdot{\bf r}_{13}\right)}\,
\frac{1}{ \omega^2_{{\bf k}_2+{\bf k}_3}\, \omega^2_{{\bf k}_2}\,
\omega^2_{{\bf k}_3}} \ ,
\end{eqnarray}
where (isospin) stands for the isospin factor in parentheses of Eq.~(\ref{eq:a11}).
After changing variables to ${\bf k}_2={\bf P}/2+{\bf p}$ and
${\bf k}_3={\bf P}/2-{\bf p}$, making use of Feynman's parametrization
for the denominator $1/\left(\omega_{{\bf P}/2+{\bf k}}\, \omega_{{\bf P}/2-{\bf k}}\right)$,
and carrying out the angular integration over the ${\bf P}$ directions, we find
\begin{equation}
I=\frac{1}{16\pi^3} \int_{-1/2}^{1/2}dy \int_0^\infty dP\, P^2 
\frac{{\rm e}^{-(P/\Lambda)^4}}{P^2+m_\pi^2} 
j_0\left(P\!\mid\! {\bf r}_1\!-\!{\bf R}_{23} +y\, {\bf r}_{23}\!\mid\right)
{\rm e}^{-L(P,y)\, r_{23}} \frac{1}{L(P,y)} \ ,
\end{equation}
where
\begin{equation}
L(P,y)=\sqrt{m_\pi^2+P^2\left(1/4-y^2\right)} \ .
\end{equation}
In terms of adimensional variables, the current now reads
\begin{eqnarray}
\label{eq:a17}
{\bf j}^{\rm N4LO}_\pm({\rm 3B},{\rm b})&=&\sum_{\tiny{\mbox{cyc}}}
\frac{g_A^3}{96\,\pi^3\, \overline{f}_\pi^4} ({\rm isospin})\,\,
{\bm \sigma}_3\cdot{\bm \nabla}^z_3\,\, {\bm \sigma}_2\cdot{\bm \nabla}^z_2\,\,
{\bm \sigma}_1\cdot{\bm \nabla}^z_1\,\, {\bm \nabla}^z_1 
 \int_{-1/2}^{1/2}\!dy \int_0^\infty\! \!dx\, x^2 \frac{{\rm e}^{-x^4}}{x^2+\overline{m}_\pi^2} \nonumber\\
&&\times \frac{{\rm e}^{-\overline{L}(x,y)\, z} } {\overline{L}(x,y)}\, j_0\left(x\!\mid\! {\bf Z} +y\, {\bf z}\!\mid\right) \ ,
\end{eqnarray}
where the gradients are relative to ${\bf z}_i=\Lambda\, {\bf r}_i$, and we have defined
${\bf Z}=\Lambda \left( {\bf r}_1-{\bf R}_{23}\right)$ and ${\bf z}=\Lambda \, {\bf r}_{23}$, and
\begin{equation}
\overline{L}(x,y)=\sqrt{\overline{m}_\pi^2+x^2\left(1/4- y^2\right)}\ .
\end{equation}
In order to evaluate the gradients, we introduce the Jacobi variables,
\begin{equation}
{\bm \nabla}^z_1={\bm \nabla}^Z\ , \qquad
{\bm \nabla}^z_2 = -\frac{1}{2}{\bm \nabla}^Z+{\bm \nabla}^z \ ,
\qquad {\bm \nabla}^z_3=-\frac{1}{2}{\bm \nabla}^Z-{\bm \nabla}^z \ ,
\end{equation}
where the gradients ${\bm \nabla}^Z$ and ${\bm \nabla}^z$ are now relative
to ${\bf Z}$ and ${\bf z}$, respectively.  We obtain
\begin{eqnarray}
&&\sigma_{3,\delta}\,\sigma_{2,\gamma}\, \sigma_{1,\beta} \left(
\frac{1}{4}\nabla^Z_\delta \nabla^Z_\gamma -\nabla^z_\delta \nabla^z_\gamma 
-\frac{1}{2}\nabla^Z_\delta \nabla^z_\gamma + \frac{1}{2}\nabla^z_\delta \nabla^Z_\gamma\right)
 \left[ {\rm e}^{-\overline{L}z}\,\,  \nabla^Z_\beta \nabla^Z_\alpha\,
 j_0\left(x\!\mid\! {\bf Z} +y\, {\bf z}\!\mid\right) \right] \nonumber\\
&=&x^2\, {\rm e}^{-\overline{L}z}\, \sigma_{3,\delta}\,\sigma_{2,\gamma}\, \sigma_{1,\beta}
\Bigg\{x^2\left(\frac{1}{4}-y^2\right) \nabla^t_\delta \nabla^t_\gamma  \nabla^t_\beta \nabla^t_\alpha 
-x\, \overline{L}\left(\frac{1}{2}-y\right) \hat{z}_\delta  \nabla^t_\gamma  \nabla^t_\beta \nabla^t_\alpha \nonumber\\
&&+x\, \overline{L}\left(\frac{1}{2}+y\right) \hat{z}_\gamma  \nabla^t_\delta  \nabla^t_\beta \nabla^t_\alpha
-\left[\overline{L}^2\, \left(1+\frac{1}{\overline{L}z}\right) \,\hat{z}_\delta\, \hat{z}_\gamma -\frac{\overline{L}}{z} 
\delta_{\gamma\delta} \right] \nabla^t_\beta \nabla^t_\alpha\Bigg\}j_0(t) \ ,
\end{eqnarray}
where we have defined ${\bf t}=x\, {\bf Z}+x\,y\, {\bf z}$ and the corresponding gradient ${\bm \nabla}^t$.
By making use of the identities
\begin{eqnarray}
 \!\!\!\!\!\nabla^t_\beta \nabla^t_\alpha\, j_0(t)&=& \delta_{\alpha\beta} \left(\frac{1}{t}\frac{d}{dt}\right) j_0(t)
 +t_\alpha\,t_\beta \left(\frac{1}{t}\frac{d}{dt}\right)^2\!\! j_0(t) \ , \\
\!\!\!\!\! \nabla^t_\gamma \nabla^t_\beta \nabla^t_\alpha\, j_0(t)&=& \left(\delta_{\alpha\beta}\, t_\gamma+
 \delta_{\alpha\gamma}\, t_\beta+\delta_{\beta\gamma}\, t_\alpha\right) \left(\frac{1}{t}\frac{d}{dt}\right)^2\!\! j_0(t)
 +t_\alpha\,t_\beta \, t_\gamma \left(\frac{1}{t}\frac{d}{dt}\right)^3\!\! j_0(t) \ , \\
 \!\!\!\!\! \nabla^t_\delta  \nabla^t_\gamma \nabla^t_\beta \nabla^t_\alpha\, j_0(t)&=&
 \left(\delta_{\alpha\beta}\, \delta_{\gamma\delta}+
 \delta_{\alpha\gamma}\, \delta_{\beta\delta}+\delta_{\beta\gamma}\, \delta_{\alpha\delta}\right) 
 \left(\frac{1}{t}\frac{d}{dt}\right)^2\!\! j_0(t) +(\delta_{\alpha\beta}\, t_\gamma\, t_\delta+
 \delta_{\alpha\gamma}\, t_\beta\, t_\delta\nonumber\\
 &&+\delta_{\beta\gamma}\, t_\alpha\, t_\delta+
 \delta_{\alpha\delta}\, t_\beta\, t_\gamma+
 \delta_{\beta\delta}\, t_\alpha\, t_\gamma+\delta_{\gamma\delta}\, t_\alpha\, t_\beta) 
 \left(\frac{1}{t}\frac{d}{dt}\right)^3 \!\!j_0(t) \nonumber\\
 &&+t_\alpha\,t_\beta \, t_\gamma\, t_\delta \left(\frac{1}{t}\frac{d}{dt}\right)^4\!\! j_0(t) \ ,
\end{eqnarray}
and
\begin{equation}
\left(\frac{1}{t}\frac{d}{dt}\right)^m\!\! j_0(t)=(-)^m \frac{1}{t^m}\, j_m(t) \ ,
\end{equation}
the current in Eq.~(\ref{eq:a17}) is reduced to a sum of terms depending on parametric
integrals in $x$ and $y$.  While the matrix element of ${\bf j}^{\rm N4LO}({\rm 3B,b})$
could in principle be evaluated, the computational effort required to do so in the present
Monte Carlo calculations is, however, too large (and unjustified in view of its expected contribution,
see Table~\ref{tb:tb1}).  For this reason it has been neglected in the present study.

\end{document}